\documentstyle{article}
\begin{document}
\thispagestyle{empty}
\begin{center}{ASPECTS OF COSMOLOGICAL RELATIVITY}\end{center}
\vspace{1cm}
\begin{center}{Moshe Carmeli}\end{center} 
\begin{center}{Department of Physics, Ben Gurion University, Beer Sheva 84105, 
Israel}\end{center}
\begin{center}{(E-mail: carmelim@bgumail.bgu.ac.il)}\end{center}
\vspace{1cm}
\begin{center}{ABSTRACT}\end{center}
In this paper we review the {\it cosmological relativity}, a new
special theory of relativity that was recently developed for cosmology, and 
discuss in detail some of its aspects. 
We recall that in this theory it is assumed that gravitation is negligible. 
Under this
assumption, the receding velocities of galaxies and the distances between them
in the Hubble expansion are united into a four-dimensional pseudo-Euclidean
manifold, similarly to space and time in ordinary special relativity. The
Hubble law is assumed and is written in an invariant way that enables one to
derive a four-dimensional transformation which is similar to the Lorentz 
transformation. The parameter in the new transformation is the ratio between 
the cosmic time to the Hubble time (in which the cosmic time is measured 
backward with respect to the present time). Accordingly, the new 
transformation relates physical quantities at different cosmic times in the
limit of weak or negligible gravitation. 

The transformation is then applied to         
the problem of the expansion of the Universe at the very early stage when 
gravity was negligible and thus the transformation is applicable. We calculate
the ratio of the volumes of the Universe at two different times $T_1$ and 
$T_2$ after the Big Bang. Under the assumptions that $T_2-T_1\approx 10^{-32}$
sec and $T_2\ll 1$ sec, we find that $V_2/V_1=10^{-16}/\sqrt{T_1}$. For $T_1
\approx 10^{-132}$ sec we obtain $V_2/V_1\approx 10^{50}$. This result 
conforms with the standard inflationary universe theory, but now it
is obtained without assuming that the Universe is propelled by antigravity.

New applications of the theory
are presented. This includes a new law for the decay of radioactive
materials, that was recently developed by Carmeli and Malin. The new law is a 
modification of the standard exponential formula, when cosmic times are 
considered instead of the ordinary local times. We also show that there is no
need to assume the existence of galaxy dark matter; the Tully-Fisher law is 
derived from our theory. A significant extension of the
theory to cosmology that was recently made by Krori, Pathak, Das and 
Purkayastha is given. In
this way cosmological relativity becomes a general theory of relativity in
seven dimensions of curved space-time-velocity. The solutions of the field 
equations in seven dimensions obtained by Krori {\it et al}. are given and 
compared to 
those of the standard Friedmann-Robertson-Walker. A completely new picture 
of the expanding Universe is thus obtained and compared to the FRW one.       
\section{Introduction} 
Few important problems in cosmology are widely discussed these days. The first 
is the problem of dark matter, its theory and its experimental verification.
This problem is intimately related to the amount of matter in the Universe, or
more accurately to the value of $\Omega=\rho/\rho_c$ where $\rho_c$ is the 
critical matter density and $\rho$ is the actual matter density in the 
Universe. 

A second problem is that of the inflation of the Universe at the 
very early stage, at which time the Universe expanded drastically. This 
problem is related to particle physics. What are the reasons for the 
inflation? Was there a kind of antigravity? A third problem is the age of the
Universe. If one determines the age of the Universe by nuclear synthesis
measurements of the Earth or our galaxy, and compares it with that obtained
from measurements of the Hubble constant (using a certain model for the 
Universe), these two ages are not exactly equal. Also the problem of directly
measuring the Hubble constant seems to depend on the distance scale of the 
galaxies used for the measurements.

In this lecture we address ourselves to the problem of the inflation at the
early stage of the Universe. At that time gravitation was in no existance. 
Within this assumption of negligible gravitation we develop a theory which 
enables us to discuss and obtain some exact results that standard methods are
unable to provide. 

Some new applications of the theory
are presented. This includes a new law for the decay of radioactive
materials, that was recently developed by Carmeli and Malin. The new law is a 
modification of the standard exponential formula, when cosmic times are 
considered instead of the ordinary local times. We also discuss the problem of
galaxy dark matter. We show that there is no need to assume the existence of
dark matter for galaxies. It is shown that the Tully-Fisher law can be derived 
from our theory. A significant extension of the
theory to cosmology that was recently made by Krori {\it et al.} is given. In
this way cosmological relativity becomes a general theory of relativity in
seven dimensions of curved space-time-velocity. The solutions of the field 
equations in seven dimensions obtained by Krori are given and compared to 
those of the standard Friedmann-Robertson-Walker. A completely new picture 
of the expanding Universe is thus obtained and compared to the FRW one.       
\section{Consequences of the Hubble Expansion}
The Hubble law expresses the simple relationship between the receding 
velocities of galaxies to their distances, thus gives a mathematical 
expression to the observation that the Universe is expanding. This is an 
experimental fact, in which underlies the assumption that the observed 
redshift in the spectrum emitted from galaxies is due to Doppler effect. In
mathematical terms the Hubble law is given by 
$$\mbox{\bf v}=H_0\mbox{\bf R}.\eqno(1)$$
$H_0$ is the Hubble constant. In reality $H_0$ is not a constant in cosmic
times; and that is due to gravity whose effect varies as the Universe expands.
In the limiting case of negligible gravitation assumed in this lecture, $H_0$ 
can be considered to be a constant and does not depend on the cosmic time. 

If we denote by $\tau=H_0^{-1}$ the Hubble time, then $\tau$ can be considered as
the age of the Universe in this particular case of neglecting gravity. We 
write the Hubble law in the trivially different form 
$$\mbox{\bf R}=\tau\mbox{\bf v},\eqno(2)$$ 
in order to compare it with the well-known expression for the propagation of
light, $R=ct$. In Eqs. (1) and (2) $\mbox{\bf R}=\left(x,y,z\right)$. Equation
(2) can thus be expressed as 
$$x^2+y^2+z^2=\tau^2v^2,\eqno(3)$$
where {\bf v} is the outgoing velocity. Accordingly
$$x^2+y^2+z^2-\tau^2v^2=0.\eqno(4)$$  

It will furthermore be assumed that a relationship of the form (4) holds at 
any cosmic time $t$. Accordingly, if we denote distances and velocities at two
different cosmic times $t$ and $t'$ by $x,y,z,v$ and $x',y',z',v'$, then it 
will be assumed that 
$$x'^2+y'^2+z'^2-\tau^2v'^2=x^2+y^2+z^2-\tau^2v^2.\eqno(5)$$  
Equation (5) resambles that for the propagation of light viewed from two 
different inertial frames of references moving with a constant velocity with 
respect to each other. In our case we have what might be called cosmic frames
of references which differ from each other by a cosmic time. 
\section{The Cosmological Transformation}
The question now arises as to what is the transformation between the four 
variables $x,y,z,v$ and $x',y',z',v'$ that leaves unaffected the invariance
equation (5)? 

For simplicity it will be assumed that $y'=y,z'=z,$ thus we have
$$x'^2-\tau^2v'^2=x^2-\tau^2v^2.\eqno(6)$$
What transformation keeps the last formula invariant? The solution of Eq. (6)
can be written as 
$$x'=x\cosh\psi-\tau v\sinh\psi,\hspace{8mm}
\tau v'=\tau v\cosh\psi-x\sinh\psi.$$
At $x'=0$ we have $\tanh\psi=x/\tau v=t/\tau.$
As a result we have 
$$\sinh\psi=\frac{t/\tau}{\sqrt{1-
\displaystyle\normalsize\frac{t^2}{\tau^2}}},\hspace{5mm}
\cosh\psi=\frac{1}{\sqrt{1-\displaystyle\normalsize\frac{t^2}{\tau^2}}}.$$
Consequently the transformation is given by 
$$x'=\frac{x-tv}{\sqrt{1-\displaystyle\normalsize\frac{t^2}{\tau^2}}},
\hspace{5mm}
v'=\frac{v-\displaystyle\normalsize\frac{tz}{\tau^2}}
{\sqrt{1-\displaystyle\normalsize\frac{t^2}{\tau^2}}},\hspace{5mm}
y'=y,z'=z.\eqno(7)$$
Here $t$ is the cosmic time measured with respect to us, now, and goes 
backward. The transformation (7) is called {\it the cosmological 
transformation}.
\section{Cosmological Special Relativity}
The transformation (7) very much resembles the well-known Lorentz 
transformation. In fact, one can give a formal foundation to establish a {\it 
cosmological special relativity} of the four-dimensional continuum of the 
three-dimensional Euclidean space and the outgoing radial velocity. We here 
mention only two consequences of the cosmological transformation, and for more 
applications and further details of such a theory the reader is referred to 
the author's book and earlier papers [1-7].
\subsection{The Law of Addition of Cosmic Times}
As is accepted nowadays, intervals of cosmic times can be added linearly. The
cosmological transformation (7), however, tells us a different thing. A
cosmological event that occured at the cosmic time $t_1$ (measured backward
with respect to us) that preceded by a second event which occured before the
first one at a cosmic time $t_2$, then the second event would appear to occur
with respect to us at a backward time $t_{12}$ given by 
$$t_{12}=\frac{t_1+t_2}{1+\displaystyle\normalsize\frac{t_1t_2}{\tau^2}}
\hspace{3mm}(cosmic\mbox{ }times\mbox{ }addition\mbox{ }law).\eqno(8)$$  
This law of addition of cosmic times can be tested by applying it to the decay
cosmic times occured in the radioactive materials in determining the ages of 
our Earth and our galaxy.
\subsection{Inflation at the Early Universe} 
At the early Universe gravity was completely negligible, and thus the 
cosmological transformation (7) may be applied. To this end we proceed as
follows.

The line element for the Universe is given by 
$$\tau^2dv^2-\left(dx^2+dy^2+dz^2\right)=ds^2.\eqno(9)$$
Hence one has
$$\tau^2\left(\frac{dv}{ds}\right)^2-\left[\left(\frac{dx}{dv}\right)^2+
\left(\frac{dy}{dv}\right)^2+\left(\frac{dz}{dv}\right)^2\right]
\left(\frac{dv}{ds}\right)^2=1,\eqno(10)$$
or
$$\left(\tau^2-t^2\right)\left(\frac{dv}{ds}\right)^2=1.\eqno(11)$$

Multiplying the last equation by $\rho_0^2$, where $\rho_0$ is the matter 
density of the Universe at the present time, we obtain for the matter density
at a backward cosmic time $t$
$$\rho=\tau\rho_0\frac{dv}{ds}=\frac{\rho_0}{\sqrt{1-\displaystyle\normalsize
\frac{t^2}{\tau^2}}}.\eqno(12)$$

Since the volume of the Universe is inversely proportional to its density, it
follows that the ratio of the volumes at two cosmic times $t_1$ and $t_2$ with
respect to us (we choose $t_2<t_1$) is given by 
$$\frac{V_2}{V_1}=\sqrt{\frac{1-\displaystyle\normalsize\frac{t_2^2}{\tau^2}}
{1-\displaystyle\normalsize\frac{t_1^2}{\tau^2}}}=
\sqrt{\frac{\left(\tau-t_2\right)\left(\tau+t_2\right)}{\left(\tau-t_1\right)
\left(\tau+t_1\right)}}.\eqno(13)$$

For cosmic times $t_1$ and $t_2$ very close to the Hubble time $\tau$, we may 
assume that $\tau+t_2\approx\tau+t_1\approx 2\tau$. Accordingly
$$\frac{V_2}{V_1}\approx\sqrt{\frac{\tau-t_2}{\tau-t_1}}.\eqno(14)$$
Denoting now by $T_1=\tau-t_1$ and $T_2=\tau-t_2$, with $T_2>T_1$. $T_1$ and
$T_2$ are the cosmic times as measured from the Big Bang. We thus have 
$$\frac{V_2}{V_1}\approx\sqrt{\frac{T_2}{T_1}}.\eqno(15)$$
For $T_2-T_1\approx 10^{-32}$ sec and $T_2\ll 1$ sec, we have
$$\frac{V_2}{V_1}\approx\sqrt{\frac{T_2}{T_1}}\approx\sqrt{\frac{T_1+10^{-32}}
{T_1}}=\sqrt{1+\frac{10^{-32}}{T_1}}\approx\frac{10^{-16}}{\sqrt{T_1}}.
\eqno(16)$$
For $T_1\approx 10^{-132}$ sec we obtain 
$$\frac{V_2}{V_1}\approx\frac{10^{-16}}{10^{-66}}=10^{50}.\eqno(17)$$

This result conforms with the inflationary universe theory of Guth [8] and 
Linde [9] without assuming
any model (such as the Universe is propelled by antigravity). 
\section{Decay Law of Radioactive Material in Cosmology}
In this section we derive, following Carmeli and Malin [10], a new cosmological
law for the decay of radioactive material. 

We assume that the probability of disintegration during any interval of {\it
cosmic} time $dt'$ is a constant. Thus 
$$\frac{dN}{dt'}=-\frac{1}{T}N,\eqno(18)$$
where $T$ is the lifetime of the material. (Throughout this section the time 
parameters $t$ and $t'$ will not be backward as are considered in the previous 
sections.) 

Now, let us substitute in the formula for the addition of cosmic times, Eq.(8),
$t_1=t=t'$ (present time), $t_2=dt$, then $t_{1+2}=t+dt'$, and making an 
approximation, using the fact that $tdt/\tau^2$ is much smaller than 1, we 
obtain 
$$t+dt'=(t+dt)(1-\frac{tdt}{\tau^2})=t+(1-\frac{t^2}{\tau^2})dt.\eqno(19)$$ 

If $dt$ is a time interval measured by a clock, and we want to obtain $dN/dt$,
we need to find $dt'/dt$ from Eq.(19) and substitute it in Eq. (18). We then 
obtain 
$$\frac{dt'}{dt}=1-\frac{t^2}{\tau^2}.\eqno(20)$$
Equations (18) and (20) subsequently yield
$$\frac{dN}{dt'}=\frac{dN}{dt}\frac{dt}{dt'}=-\frac{1}{T}N,\eqno(21)$$
or, using Eq.(20),
$$\frac{dN}{N}=-\frac{1}{T}\frac{dt'}{dt}dt=-\frac{1}{T}\left(1-
\frac{t^2}{\tau^2}\right)dt.\eqno(22)$$
Integration of the last equation then gives 
$$N(t)=N_0\exp\left[-\frac{1}{T}\left(1-\frac{t^2}{3\tau^2}\right)t\right].
\eqno(23)$$ 

Now, when we say that $t$ and $t'$ are present time, we start them at $t=t'=
0$, which is the time when $N=N_0$. Eq.(23) will provide large deviations from
Eq.(18) when $T$ is comparable to $\tau$, and we measure radioactivity over
astronomical times. It is not clear how such measurements/observations can be
carried out. However, it may be possible to detect minute deviations from 
linearity in a graph of $\ln N$ vs. $t$ in very accurate laboratory 
measurements.

In principle, it follows from Eq.(23) that $N(t)$ for a given $t$ is less than
the traditional formula predicts. Namely, the material decays faster that
expected.
\section{Galaxy Dark Matter as a Property of Spacetime}
In this section we generalize cosmological relativity to curved space. This
will enable us to introduce gravitation. 

We first describe, following Carmeli
[11], the motion of a star in a central field of
a galaxy in an expanding universe. Use is made of a double expansion in $1/c$
and $1/\tau$. In the lowest approximation the rotational velocity of the star
will be shown to satisfy $v^4=\frac{2}{3}GMcH_0$, where $G$ is Newton's 
gravitational constant and $M$ is the mass of the galaxy. This formula 
satisfies
observations of stars moving in spiral and elliptical galaxies, and is in 
accordance with the Tully-Fisher law [12,13].

The problem of motion in general relativity is a very old one and started with 
Einstein and Grommer [14] who showed that the equations of motion follow from the 
Einstein field equations rather than have to be postulated independently as in
electrodynamics. This is a consequence of the nonlinearity of the field 
equations and the Bianchi identities. Much work was done since then and the 
problem of motion in the gravitational field of an isolated system is well
understood these days [15-26].

The topic of motion in an expanding universe is of considerable importance in
astronomy since stars moving in spiral and elliptical galaxies show serious
deviation from Newtonian gravity and the latter follows from general relativity 
theory [27]. It follows that the Hubble expansion imposes an extra constraint
on the motion -- the usual assumptions made in deriving Newtonian gravity from 
general relativity are not sufficient in an expanding universe. The star is 
not isolated from the ``flow" of matter in the universe. When this is taken 
into account, along with Newton's gravity, the result is a motion which 
satisfies a different law from the one determining the planetary motion in the
solar system.
\subsection{Geodesic Equation}
The equation that describes the motion of a simple particle is the geodesic
equation. It is a direct result of the Einstein field equation $G_{\mu\nu}=
\kappa T_{\mu\nu}(\kappa=8\pi G/c^4)$. The restricted Bianchi identities 
$\nabla_\nu G^{\mu\nu}\equiv 0$ implies the covariant conservation law 
$\nabla_\nu T^{\mu\nu}=0$. When volume-integrated, the latter yields the 
geodesic equation. To obtain the Newtonian gravity it is sufficient to assume
the approximate forms for the metric: $g_{00}=1+2\phi/c^2$, $g_{0k}=0$ and
$g_{kl}=-\delta_{kl}$, where $k,l=1,2,3,$ and $\phi$ a function that is 
determined by the Einstein field equations. In the lowest approximation in 
$1/c$ one then has
$$\frac{d^2x^k}{dt^2}=-\frac{\partial\phi}{\partial x^k},\eqno(24)$$
$$\nabla^2\phi=4\pi G\rho,\eqno(25)$$
where $\rho$ is the mass density. For a central body $M$ one then has $\phi=
-GM/R$ and Eq.(25) yields, for circular motion, the first integral 
$$v^2=GM/R,\eqno(26)$$
where $v$ is the rotational velocity of the particle.
\subsection{Hubble's Law in Curved Space}
The Hubble law was given by Eq.(1) and recasted in the form of Eq.(4) when 
gravity was neglected. Gravitation, 
however, does not permit global linear relations like Eq.(4) and the latter 
has to be adopted to curved space. To this end one has to modify Eq.(4) to the
differential form and to adjust it to curved space. The generalization of 
Eq.(4) is, accordingly,
$$ds^2=g'_{\mu\nu}dx^\mu dx^\nu=0,\eqno(27)$$
with $x^0=\tau v$. Since the universe expands radially (it is assumed to be
homogeneous and isotropic), it is convenient to use spherical coordinates
$x^k=(R,\theta,\phi)$ and thus $d\theta=d\phi=0$. We are still entitled to
adopt coordinate conditions, which we choose as $g'_{0k}=0$ and $g'_{11}=
g'_{00}{}^{-1}$. Equation (5) reduces to
$$\frac{dR}{dv}=\tau g'_{00}.\eqno(28)$$
This is Hubble's law taking into account gravitation, and hence dilation and 
curvature. When gravity is negligible, $g'_{00}\approx 1$ thus $dR/dv=\tau$ 
and by integration, $R=\tau v$ or $v=H_0R$ when the initial conditions are 
chosen appropriately.
\subsection{Phase Space}
As is seen, the Hubble expansion causes constraints on the structure of the
universe which is expressed in the phase space of distances and velocities,
exactly the observables. The question arises: What field equations the metric
tensor $g'_{\mu\nu}$ satisfies? We {\it postulate} that $g'_{\mu\nu}$ 
satisfies the Einstein field equations in the phase space, $G'_{\mu\nu}=K
T'_{\mu\nu}$, with $K=8\pi k/\tau^4$, and $k=G\tau^2/c^2$. Accordingly, in
cosmology one has to work in both the real space and in the phase space. 
Particles follow geodesics of both spaces (in both cases they are consequences
of the Bianchi identities). For a spherical solution in the phase space, 
similarly to the situation in the real space, we have in the lowest 
approximation in $1/\tau$ the following: $g'_{00}=1+2\psi/\tau^2$, $g'_{0k}=0$
and $g'_{kl}=-\delta_{kl}$, wuth $\nabla^2\psi=4\pi k\rho$. For a spherical 
solution we have $\psi=-kM/R$ and the geodesic equation yields
$$\frac{d^2x^k}{dv^2}=-\frac{\partial\psi}{\partial x^k},\eqno(29)$$  
with the first integral 
$$\left(\frac{dR}{dv}\right)^2=\frac{kM}{R},\eqno(30)$$
for a rotational motion. Integration of Eq.(30) then gives
$$R=\left(\frac{3}{2}\right)^{2/3}\left(kM\right)^{1/3}v^{2/3}.\eqno(31)$$
Inserting this value of $R$ in Eq.(26) we obtain
$$v^4=\frac{2}{3}GMcH_0.\eqno(32)$$
\subsection{Galaxy Dark Matter}
The equation of motion (32) has a direct relevance to the problem of the
existence of the galaxy dark matter. As is well known, observations show that
the fourth power of the rotational velocity of stars in some galaxies is
proportional to the luminousity of the galaxy (Tully-Fisher law), $v^4\propto
L$. Since the luminousity, by turn, is proportional to the mass $M$ of the
galaxy, $L\propto M$, it follows that $v^4\propto M$, independent of the 
radial distance of the star from the center of the galaxy, and in violation to
Newtonian gravity. Here came the idea of galaxy dark matter or, alternatively,
modification of Newton's gravity in an expanding universe.

We have seen how a careful application of general relativity theory and 
cosmological relativity give an
answer to the problem of motion of stars in galaxies in an expanding universe.
If Einstein's general relativity theory is valid, then it appears that the 
galaxy halo dark matter is a property of spacetime and not some physical 
material. The situation resembles that existed at the beginning of the century
with respect to the problem of the advance of the perihelion of the planet
Mercury which general relativity showed that it was a property of spacetime
(curvature).
\section{Carmeli's Cosmology}
Based on the full group of transformations of cosmological relativity in the
seven-dimensional space of space-time-velocity, Krori, Pathak, Das and 
Purkayastha [28] have extended the flat-space metric to describe what the
authors call Carmeli's cosmology for the expanding
universe. The properties of this new cosmology were discussed in detail and
compared with the standard FRW cosmology. The following is based on the paper
by Krori {\it et al}, and for the full details the reader is referred to the
original paper.
\subsection{Carmeli's Cosmological Metric}
The starting point was the flat-space Carmeli's metric 
$$ds^2=c^2dt^2-(dx^2+dy^2+dz^2)+\tau^2(dv_x^2+dv_y^2+dv_z^2).\eqno(33)$$
The metric (33) was subsequently extended to the following form:
$$ds^2=c^2dt^2-R^2(t)(dx_1^2+dx_2^2+dx_3^2)+T^2(t)(dv_1^2+dv_2^2+dv_3^2).
\eqno(34)$$ 
Here $R(t)$ is the three-space scale factor and $T(t)=H^{-1}(t)$, where 
$H(t)$ is the Hubble parameter.
\subsection{Field Equations}
The energy-momentum tensor components are given by
$$T^0_{\hspace{6pt}0}=\rho c^2,\hspace{5mm} T^1_{\hspace{6pt}1}=
T^2_{\hspace{6pt}2}=T^3_{\hspace{6pt}3}=-p,\hspace{5mm}\mbox{\rm and}
\mbox{ }T^\mu_{\hspace{6pt}\nu}=0;\mbox{ }\mu,\nu\geq 4.\eqno(35)$$ 
From Eqs.(34) and (35), using Einstein's field equations in seven dimensions,
we obtain 
$$\frac{\dot{R}^2}{R^2}+\frac{\dot{T}^2}{T}+\frac{3\dot{R}\dot{T}}{RT}=
\frac{8\pi c^4\rho}{3},\eqno(36)$$
$$\frac{2\ddot{R}}{R}+\frac{\dot{R}^2}{R^2}+\frac{3\ddot{T}}{T}+
\frac{3\dot{T}^2}{T^2}+\frac{6\dot{R}\dot{T}}{RT}=-8\pi c^2p,\eqno(37)$$
$$\frac{3\ddot{R}}{R}+\frac{3\dot{R}^2}{R^2}+\frac{2\ddot{T}}{T}+
\frac{\dot{T}^2}{T^2}+\frac{6\dot{R}\dot{T}}{RT}=0,\eqno(38)$$
where a dot denotes differentiation with $t$. 
We thus have three equations for the four unknown variables $R$, $T$, $\rho$ 
and $p$. 
\subsection{Solution of the Field Equations}
One assumes a solution of the form $R=R_0t^m$, where $m$ is a 
positive parameter, and putting $\dot{T}=Tu(t)$, we obtain
$$t^2\dot{u}+\frac{3}{2}t^2u^2+3mtu+\frac{3}{2}m(2m-1)=0.\eqno(39)$$
We next define a function $v(t)$ such that $2\dot{v}-3uv=0$, Eq.(39) will
then have the form
$$t^2\ddot{v}+3mt\dot{v}+\frac{9}{4}m(2m-1)v=0.\eqno(40)$$

Equation (40) admits a solution of the form $v=v_0t^n$, where $v_0$ is a 
constant and $n$ is given by
$$n=\frac{1-3m\pm\sqrt{3m+1-9m^2}}{2}.\eqno(41)$$
From the above we obtain $u=2n/3t$ and thus $T=T_0t^{2n/3}$, where $T_0$ is a
constant. Now, using the expressions of $R$ and $T$ in Eqs.(36) and (37) we
finally obtain
$$8\pi c^4\rho t^2=3m^2+4n^2/3+6mn,\eqno(42)$$
$$8\pi c^2pt^2= -3m^2+2m-8n^2/3+2n-4mn.\eqno(43)$$
Equations (42) and (43) provide a complete solution of the field equations.  
\subsection{Properties of Carmeli's Cosmology}
The properties of Carmeli's cosmology are now discussed and compared with the
FRW cosmology [29]. 

(a) $m=\frac{1}{2}$ corresponds to radiation era for both the cosmologies. At 
this value of $m$, $n=0$ in Carmeli cosmology.

(b) In Carmeli cosmology, $m$ varies from $1/3$ to $(1+\sqrt{5})/6$ with $\rho
c^2/p$ correspondingly decreasing from $\infty$ (dust) to 1.6191714. On the
other hand, in FRW cosmology, $m$ varies from $1/3$ to $2/3$ with $\rho c^2/p$ 
correspondingly increasing from 1 (stiff matter) to $\infty$ (dust). Obviously 
stiff matter is not admissible in Carmeli cosmology.

(c) In Carmeli cosmology, as $m$ increases from $1/3$ to $(1+\sqrt{5})/6$, $p$
increases from 0, reaches a maximum and then falls to a certain non-zero 
value.

(d) In FRW cosmology, the Hubble parameter, $H=\dot{R}/R$, decreases with the 
passage of time for all non-zero values of $m$. On the other hand in Carmeli
cosmology, the Hubble parameter, $H=T^{-1}$, decreases with the passage of 
time for $m<1/2$ (with $n$ positive), increases with the passage of time for 
$m>1/2$ (with $n$ negative) but is constant for $m=1/2$ (with $n=0$).
 
(e) In FRW cosmology, in the present (dust) era ($m=2/3$), the Hubble parameter 
$H_{FRW}$ and the age of the Universe $t_{FRW}$ are related by the formula
$$H_{FRW}=\frac{2}{3}t_{FRW}^{-1}.$$

In Carmeli's cosmology, since $T=H^{-1}$ the Hubble parameter $H_c$ and the 
age of the present (dust) universe, $t_c$, are related by 
$$H_c=\frac{1}{T_0}t^{-2n/3}.$$

Following FRW cosmology, if we take $1/T_0$ to be a factor of order 1, then 
with $H_c=H_{FRW}$ in the present era, we have $t_c\sim t_{FRW}^{3/2n}$, i.e.
$t_c\sim t_{FRW}^3$, since $n=1/2$.

If this is so, then the age of the universe is very much higher according to 
Carmeli cosmology. Further, $H_c$ decreases more slowly than $H_{FRW}$ with
the passage of time in the present era.

(f) The expression for cosmological redshift is [30] 
$$1+z=\frac{R(t_0)}{R(t_1)},$$ 
where 
$t_0$ is the epoch at which light emitted at an earlier epoch $t_1$ from a 
distant galaxy is received by our galaxy. Now, if $t_0$ and $t_1$ are assumed 
to be the same in both the cosmologies, then for FRW cosmology (present era)
$$1+z_{FRW}=\left(\frac{t_0}{t_1}\right)^{2/3},$$ 
since $m=2/3$, and for Carmeli cosmology (present era)
$$1+z_c=\left(\frac{t_0}{t_1}\right)^{1/3},$$ since $m=1/3$.

Obviously, $z_{FRW}>z_c$. In other words, redshift is less red in Carmeli 
cosmology.

(g) The expression for the angular size of a distant galaxy is [30]
$$\Delta\theta=\frac{d}{r_1R\left(t_1\right)}=\frac{d\left(1+z\right)}{r_1R
\left(t_0\right)},\eqno(44)$$
where $d$ is the breadth of the distant galaxy, $r_1$ is its coordinate 
distance from our galaxy, $t_0$ is the epoch of observation (in our galaxy) of
the light emitted at the earlier epoch, $t_1$, by the distant galaxy and $z$
is the redshift. If $r_1$ and $t_1$ are assumed to be same for both 
cosmologies, then, from Eq.(44), 
$$\Delta\theta_c>\Delta\theta_{FRW},$$ 
since $m=2/3$ for FRW cosmology and $m=1/3$ for Carmeli cosmology at the 
present (dust) era.

(h) The expression for luminosity distance of a distant galaxy is 
$$D=r_1R(t_0)(1+z),$$ 
where $r_1$ is the coordinate distance of the distant galaxy and $t_0$
is the epoch at which light is received by our galaxy from the distant galaxy.
Since, as already seen, above, $$R_{FRW}(t_0)>R_c(t_0)$$ and $$z_{FRW}>z_c.$$ 
We find that $$D_{FRW}>D_c.$$

It may be noted in time that Carmeli cosmology is theoretically interesting
in its own right. However, in the context of its significant deviations from 
the standard FRW cosmology, sophisticated observational techniques will have 
to be devised to assess the worth and validity of this new cosmology.
\section{Appendix: Table of Numerical Results}
We list below numerical results as given by Krori {\it et al.} [28].\vspace{1mm}\newline

\begin{tabular}{ccccc}
$m$&$n$&$8\pi\rho c^4t^2$&$8\pi pc^2t^2$&$\rho c^2/p$\\ \hline\hline
1/3&1/2&5/3&0&$\infty$\\0.4&0.3358898&1.4365645&0.1534974&9.3588849\\
0.458&0.1616129&1.1082291&0.244209&4.5380355\\
0.474&0.1051945&0.9879554&0.2554032&3.8682186\\
1/2&0&0.7500000&0.2500000&3.000000\\
0.514&-0.0543697&0.6288532&0.2305139&2.7280484\\
0.53&-0.1706014&0.3389937&0.161592&2.0978371\\
$\left(1+\sqrt{5}\right)/6$&$-\left(\sqrt{5}-1\right)/4$&0.00099664&
0.000615572&1.6191714\\ \hline
\end{tabular}
\section*{References}
1. M. Carmeli, Cosmological relativity: a special relativity for cosmology, 
{\it Found. Phys.} {\bf 25}, 1029 (1995).\newline
2. M. Carmeli, Extension of the Lorentz group to cosmology, {\it Commun. 
Theor. Phys.} {\bf 4}, 109 (1995).\newline
3. M. Carmeli, Space, time and velocity, {\it Commun. Theor. Phys.} {\bf 4}, 
233 (1995).\newline
4. M. Carmeli, Cosmological special relativity, {\it Found. Phys.} {\bf 26}, 
413 (1996).\newline
5. M. Carmeli, Structure of cosmology in $\Omega=1$ universe, {\it Commun. 
Theor. Phys.} {\bf 5}, 65 (1996).\newline
6. M. Carmeli, Space, time and velocity in cosmology, {\it Intern. J. Theor. 
Phys.} {\bf 36}, 757 (1997).\newline 
7. M. Carmeli, {\it Cosmological Special Relativity: The Large-Scale 
Structure of Space, Time and Velocity}, World Scientific (1997).\newline 
8. A.H. Guth, {\it Phys. Rev. D} {\bf 23}, 347 (1981).\newline 
9. A.D. Linde, {\it Phys. Lett. B} {\bf 116}, 335 (1982).\newline
10. M. Carmeli and S. Malin, A new formula for the decay of radioactive 
material, to be published.\newline
11. M. Carmeli, Is galaxy dark matter a property of spacetime?, {\it Commun. 
Theor. Phys.} {\bf 6}, 101 (1997); {\it Intern. J. Theor. Phys.} {\bf 37}, 
No. 10 (1998).\newline
12. R.B. Tully and J.R. Fisher, {\it Astron. Ap.} {\bf 54}, 661 (1977).\newline
13. P.J.E. Peebles, {\it Principles of Physical Cosmology}, Princeton 
University Press (1993), p.49.\newline
14. A. Einstein and J. Grommer, {\it Sitz. Preuss. Acad. Wiss. Phys. Math. 
K.} {\bf 1}, 2 (1927).\newline
15. A. Einstein, L. Infeld and B. Hoffmann, {\it Ann. Math.} {\bf 39}, 65 
(1938).\newline
16. A. Einstein and L. Infeld, {\it Canad. J. Math.} {\bf 1}, 209 (1949).
\newline 
17. V. Fock, {\it The Theory of Space, Time and Gravitation}, Pergamon Press 
(1959).\newline  
18. L. Infeld and J. Plebanski, {\it Motion and Relativity}, Pergamon Press
(1960).\newline 
19. M. Carmeli, The Einstein-Infeld-Hoffmann equations of motion up to the 
ninth order, {\it Phys. Lett.} {\bf 9}, 132 (1964).\newline
20. M. Carmeli, The motion of a particle of finite mass in an external 
gravitational field, {\it Ann. Phys. (N.Y.)} {\bf 30}, 168 (1964).\newline 
21. M. Carmeli, Motion of a charge in gravitational field, {\it Phys. Rev.} 
{\bf 138}, B1003 (1965).\newline 
22. M. Carmeli, The equations of motion of slowly moving particles in the
general theory of relativity, {\it Nuovo Cimento} {\bf X37}, 842 (1965).\newline 
23. M. Carmeli, Semigenerally covariant equations of motion -- I: derivation, 
{\it Ann. Phys. (N.Y.)} {\bf 34}, 465 (1965).\newline 
24. M. Carmeli, Semigenerally covariant equations of motion -- II: the 
significance of the tail and the relation to other equations of motion, 
{\it Ann. Phys. (N.Y.)} {\bf 35}, 250 (1965).\newline 
25. M. Carmeli, Equations of motion without infinite self-action terms in 
general relativity, {\it Phys. Rev.} {\bf 140}, B1441 (1965).\newline 
26. T. Damour, The motion of compact bodies, in: {\it Gravitational Radiation}, 
N. Deruelle and T. Piran, Eds., North-Holland, 1983, pp. 59--144.\newline
27. M. Carmeli, {\it Classical Fields: General Relativity and Gauge Theory},
John Wiley (1982), Chapter 6.\newline
28. K.D. Krori, K. Pathak, Kanika Das and Anuradha Das Purkayastha, Carmeli 
cosmology, to be published.\newline 
29. H. Ohanian and R. Ruffini, {\it Gravitation and Spacetime} (Second 
Edition), W.W. Norton, New York (1994), Chapter 9.\newline
30. J.V. Narlikar, {\it Introduction to Cosmology}, Cambridge University
Press (1993), pp. 92--100.\newline
\end{document}